\documentclass[12pt]{article}
\usepackage{times}
\usepackage{codams,float,amssymb}
\usepackage[leqno]{amsmath}
\usepackage[dvipdf]{graphicx}
\floatstyle{ruled}
\newfloat{displaycode}{htp}{alg}[section]
\floatname{displaycode}{Eiffel code}
\usepackage{wrapfig}
\usepackage{noweb}
\noweboptions{breakcode}
\DeclareFontFamily{OT1}{codefont}{}
\DeclareFontShape{OT1}{codefont}{m}{n}{ <-> cmtt12 }{}

\let\Tt=\codefont

\numberwithin{equation}{section}

\DeclareGraphicsExtensions{.eps}
\title{CUDA implementation of Wagener's 2D convex
hull PRAM algorithm}
\newcommand{\hb}{\hfil\break}
\author{Colm \'O D\'unlaing\thanks{e-mail: odunlain@maths.tcd.ie.
Mathematics department website: http://www.maths.tcd.ie.}\\
{\em Mathematics, Trinity College, Dublin 2, Ireland}}

\begin{document}
\pagestyle{noweb}
\maketitle
\begin{abstract}
This paper describes a CUDA implementation of Wagener's
PRAM convex hull algorithm in $\IR^2$
[\ref{wagener},\ref{od}]. It is presented
in Knuth's literate programming style.
\end{abstract}

\section{Using this file}
\label{sect: using this file}

The source of this document is a .nw file (for
`noweb,' an implementation of Knuth's literate
programming technique: see `Literate programming
with noweb,' by Andrew L. Johnson and Brad C. Johnson,
{\em Linux Journal,} October 1st 1997).
Noweb allows one to mix LaTeX with C (or pretty
well any programming language), allowing a well-annotated
program.  One can extract `chunks' from it.
You need the noweb system, of course (that is,
notangle to extract the C part and noweave
to typeset the full document).

This document includes a Makefile.  To start
the ball rolling, you  can extract it as follows:

\beginvbox
notangle -t8 -Rwagener.Makefile wagener.nw > wagener.Makefile
\endvbox

With it, you can make a CUDA source file (wagener.cu)
or a DVI copy of this document (make dvi produces wagener.dvi)

There is one problem with wagener.cu.  The construct
{\Tt{}<<<\nwendquote}\ldots{\Tt{}>>>\nwendquote} is a necessary part of the cuda source
code, and it conflicts with noweb's
construct {\Tt{}<<\nwendquote}\ldots{\Tt{}>>\nwendquote}.  Therefore wagener.cu contains

\begin{center}
{\Tt{}match{\_}and{\_}merge\ \ LLL\ range,\ block\ RRR\ (\ hood,\ newhood,\ scratch\ );\nwendquote}
\end{center}

\noindent and it must be edited, changing
{\Tt{}LLL\nwendquote} to {\Tt{}<<<\nwendquote} and {\Tt{}RRR\nwendquote} to {\Tt{}>>>\nwendquote}.

\nwfilename{wagener.nw}\nwbegincode{1}\moddef{copyright}\endmoddef\nwstartdeflinemarkup\nwenddeflinemarkup
/*
 *  Copyright (C) 2010-12 Colm O Dunlaing (odunlain@maths.tcd.ie)
 *
 *  This file is free software: you can redistribute it and/or modify
 *  it under the terms of the GNU General Public License as published
 *  by the Free Software Foundation, either version 3 of the License, or
 *  (at your option) any later version.
 *
 *  This program is distributed in the hope that it will be useful,
 *  but WITHOUT ANY WARRANTY; without even the implied warranty of
 *  MERCHANTABILITY or FITNESS FOR A PARTICULAR PURPOSE.  See the
 *  GNU General Public License for more details.
 *
 *  You should have received a copy of the GNU General Public License
 *  along with this program.  If not, see <http://www.gnu.org/licenses/>.
*/

\nwendcode{}\nwbegindocs{2}\nwdocspar

\section{Wagener's algorithm, CUDA version}

The present goal is to make a working CUDA version
of Wagener's PRAM algorithm for computing an
upper hood for a set of $n$ points presented
in left-to-right order.

We have not considered the memory access patterns
which may seriously degrade performance.  Again,
thread divergence may degrade performance --- it
is an interesting exercise to write `non-divergent'
code.  This has been done in some places and not
in others.

This program assumes that 
\begin{itemize}
\item $n$ (the number of points)
is a power of 2.
\item No three points are collinear.
\item
All $x$-coordinates are between 0 and 1.
We shall use the point {\tt REMOTE}, $(10,0)$,
for padding.  Any point whose $x$-coordinate
is $>1$ is assumed to be `remote,' used for padding.
\item There are no floating-point errors (i.e.,
it's a problem, but it's not our problem.)
\end{itemize}

Also, three shared device arrays, one {\tt short} and
two {\tt float2} are used, of size $n$. Their total size is $18n$ bytes.
This puts
inessential limitations on $n$ --- there would be
no difficulty, and little overhead, in slicing the data into manipulable
chunks for larger $n$.

\nwenddocs{}\nwbegincode{3}\moddef{wagener}\endmoddef\nwstartdeflinemarkup\nwenddeflinemarkup
\LA{}copyright\RA{}
\LA{}globals\RA{}
\LA{}match and merge\RA{}
\LA{}main\RA{}

\nwendcode{}\nwbegincode{4}\moddef{globals}\endmoddef\nwstartdeflinemarkup\nwenddeflinemarkup
#include <stdio.h>
#include <stdlib.h>
#include <cuda.h>

float2 * point;
int count;
float2 * host_hood;
        /* The following are device variables */
float2 * hood, * newhood; short * scratch;

float2 REMOTE = \{ 10.0f, 0.0f \};

        /***************************************
         * make_remote () without memcopy
         ***************************************/

__device__ void make_remote ( float2 * p )
\{
  p->x = 10.0f; p->y = 0.0f;
\}

\nwendcode{}\nwbegindocs{5}\nwdocspar

Points are stored in the array {\tt point}, and initially
copied to {\tt host\_hood}.  The main program launches
the {\tt global} routine {\tt match\_and\_merge} repeatedly
to merge adjacent hoods from intervals of size $d$
to hoods of size $2d$.

The algorithm repeatedly copies {\tt host\_hood[]} to
device array {\tt hood[]}, launches\hb
{\tt match\_and\_merge()},
and copies the device array {\tt newhood[]} to {\tt host\_hood[]}.

\begin{figure}
\centerline{
\includegraphics[height=2in]{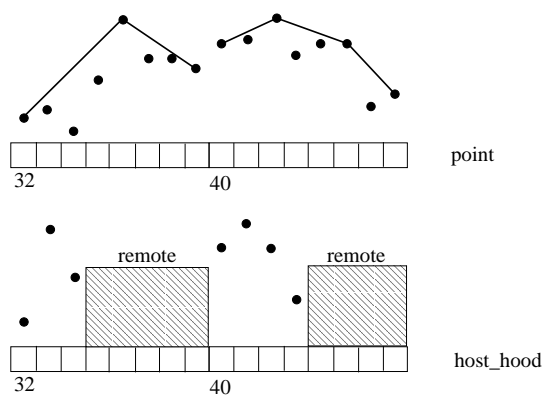}
}
\caption{Points and hoods.  The $x$-coordinates have
been distorted in the depiction of {\tt host\_hood}.}
\label{fig: host_hood}
\end{figure}

Let $s = \log_2 n$; $s$ is a positive integer.
The hood is built in $s-1$ stages (there is nothing to do if $s=1$).
At the $r$-th stage, let $d=2^r$: {\tt host\_hood} defines $n/d$
hoods.
For $0 \leq \ell < n/d$, let $P$ be the $\ell$-th block
of $d$ points from {\tt point} (indexed from $\ell d$
to $\ell d + d - 1$). The $\ell$-th hood
is $H(P)$.  The corners of $H(P)$ are stored in the
corresponding block of {\tt host\_hood}, shifted left and
padded with copies of {\tt REMOTE}
(Figure \ref{fig: host_hood}).

Next,
$n/2$ {\tt match\_and\_merge} threads are launched in
$n/(2d)$ blocks of dimension $d_1\times d_2$, where
$d_1 = 2^{\lceil r/2\rceil}$ and
$d_2 = 2^{\lfloor r/2\rfloor}$, so $d = d_1 d_2$.
The $\ell$-th block of threads cooperate to
compute $H(P\cup Q)$, where $P$ and $Q$ are
the $2\ell$-th and $2\ell+1$-st interval of $d$ points,
locating the common
tangent of $H(P)$ and $H(Q)$ and replacing these
separate hoods by $H(P\cup Q)$, shifted and padded
in a block of $2d$ entries in {\tt hood}.

The routine {\tt make\_remote(float2 *p)} is used
to set a point to remote values (I'm not sure
how to assign a constant {\tt float2} value in device code).

\nwenddocs{}\nwbegincode{6}\moddef{main}\endmoddef\nwstartdeflinemarkup\nwenddeflinemarkup

int pos_power_of_2 ( int x )
\{
  if ( x < 2 )
    return 0;

  while ( x > 1 )
    if ( x 
      return 0;
    else if ( x == 2 )
      return 1;
    else
      x /= 2;
\}

void show_current_hoods ( FILE * outfile, int d )
\{
  int i, j, hoodsize;
  fprintf(outfile, "
  for ( i=0; i<count/d; ++i )
  \{
    hoodsize = 0;
    for ( j=0; j < d; ++j )
      if ( host_hood[i*d+j].x <= 1.0 )
        ++ hoodsize;
    fprintf(outfile,"
    for ( j=0; j<d; ++j )
      if ( host_hood[i*d+j].x <= 1.0 )
        fprintf(outfile,"
                host_hood[i*d+j].y);
  \}
  fprintf(outfile,"\\n");
\}

main( int argc, char * argv[] )
\{
  int i;
  int d, d1, d2;
  FILE * file;
  FILE * trace;
  short * h_scratch;

  count = 0;

  if ( argc != 2 && argc != 3 )
  \{
    fprintf(stderr,
        "usage: 
        argv[0]);
    exit(-1);
  \}

\nwendcode{}\nwbegindocs{7}\nwdocspar

{\sc The program copies} the points to standard output, computes
the hood and writes the hood points to standard output.
It may write comment lines beginning {\tt \#}.  If the
trace file is used, it prints the intermediate hood sequences
to this file.  The program output is intended to be
sent to a companion program {\bf hood2ps} which generates
postscript.

\nwenddocs{}\nwbegincode{8}\moddef{main}\plusendmoddef\nwstartdeflinemarkup\nwenddeflinemarkup
  file = fopen ( argv[1], "r" );
  if ( file == NULL )
  \{
    fprintf(stderr,"
    exit(-1);
  \}

  trace = NULL;
  if ( argc == 3 )
  \{
    trace = fopen ( argv[2], "w" );
    if ( trace == NULL )
      fprintf(stderr,"Can't write to 
  \}

  fscanf(file,"
  if ( ! pos_power_of_2 ( count ) )
  \{
    fprintf(stderr, "Count 
    exit(-1);
  \}

  printf ("

  point = (float2*) malloc (count * sizeof(float2) );
  host_hood = (float2*) malloc (count * sizeof(float2) );
  h_scratch = (short*) malloc ( count * sizeof ( short ) );

  for (i=0; i<count; ++i)
  \{
    fscanf(file, "
    printf("
    host_hood[i] = point[i];
  \}
  printf("\\n");

  d1 = 2;
  d2 = 1;
  d  = d1 * d2;

  hood = newhood = NULL;
  scratch = NULL;

\nwendcode{}\nwbegindocs{9}\nwdocspar
{\sc The array {\tt point} will contain the} data points,
and {\tt host\_hood}
will contain the intermediate hoods as illustrated
in Figure \ref{fig: host_hood}.
{\tt H\_scratch} is to hold a copy, on the host, of
the device array {\tt scratch}, for debugging.
The shared device arrays {\tt hood, newhood, scratch}
are allocated at every thread launch.  Also, {\tt host\_hood}
needs to be copied to {\tt hood} before the thread
launch.

\nwenddocs{}\nwbegincode{10}\moddef{main}\plusendmoddef\nwstartdeflinemarkup\nwenddeflinemarkup
  while ( d < count )
  \{
    if ( trace != NULL )
      show_current_hoods ( trace, d );
    if ( hood != NULL )
    \{
      cudaFree ( hood );
      cudaFree ( newhood );
      cudaFree (scratch);
    \}

    cudaMalloc( (void **) & hood, count * sizeof( float2 ));
    cudaMemcpy( hood, host_hood, count * sizeof(float2),
        cudaMemcpyHostToDevice);

    cudaMalloc( (void **) & newhood, sizeof( float2 ) * count );
    cudaMalloc( (void **) & scratch, count * sizeof(short) );
   
\nwendcode{}\nwbegindocs{11}\nwdocspar
\noindent
{\sc
Now the thread launch:} $n$ threads in $n/(2d)$ blocks
of dimension $d_1\times d_2$.

\nwenddocs{}\nwbegincode{12}\moddef{main}\plusendmoddef\nwstartdeflinemarkup\nwenddeflinemarkup

        /*
         * LLL and RRR need to be replaced
         * by triple < and >: double < and >
         * have a special meaning in noweb,
         * the literate programming system
         * we use.
         */

    dim3 range ( count / (2*d) );
    dim3 block ( d1, d2 );
    match_and_merge  LLL range, block RRR ( hood, newhood, scratch );

\nwendcode{}\nwbegindocs{13}\nwdocspar
\noindent
{\sc When all threads have terminated,}
copy the revised array {\tt newhood} to {\tt host\_hood},
and print various debugging items.

\nwenddocs{}\nwbegincode{14}\moddef{main}\plusendmoddef\nwstartdeflinemarkup\nwenddeflinemarkup

    cudaMemcpy(host_hood, newhood, count * sizeof(float2),
        cudaMemcpyDeviceToHost);

    printf("#returned from match_and_merge, d1=
        d1, d2, d);

    cudaMemcpy(h_scratch, scratch, count * sizeof (short),
        cudaMemcpyDeviceToHost);
\nwendcode{}\nwbegindocs{15}\nwdocspar
\newpage
\nwenddocs{}\nwbegincode{16}\moddef{main}\plusendmoddef\nwstartdeflinemarkup\nwenddeflinemarkup
    printf("#scratch:\\n#");
    for(i=0; i<count; ++i )
    \{
      printf("
      if ( i > 0 && i 
        printf("\\n#");
    \}
    printf("\\n");

    if ( d1 > d2 )
      d2 *= 2;
    else
      d1 *= 2;
    d = d1 * d2;
  \}

\nwendcode{}\nwbegindocs{17}\nwdocspar
The following is for debugging.

\nwenddocs{}\nwbegincode{18}\moddef{main}\plusendmoddef\nwstartdeflinemarkup\nwenddeflinemarkup

  cudaMemcpy(host_hood, newhood, count * sizeof(float2),
        cudaMemcpyDeviceToHost);

  printf("#newhood contents\\n");
  for (i=0; i<count; ++i)
    printf("#

  if ( trace != NULL )
  \{
    fprintf(trace,"0\\n");
    fclose ( trace );
  \}

  show_current_hoods ( stdout, count );

  return 0;
\}

\nwendcode{}\nwbegindocs{19}\nwdocspar

\begin{minipage}{3in}
The remaining functions are on the device.
The function {\tt left\_of()} returns 1 if $r$ is left
of the directed line-segment $pq$,\\
(i.e., $\det(q-p,r-p)>0$), 0 otherwise.
\end{minipage}
\hfil
\begin{minipage}{2in}
\includegraphics[width=1.8in]{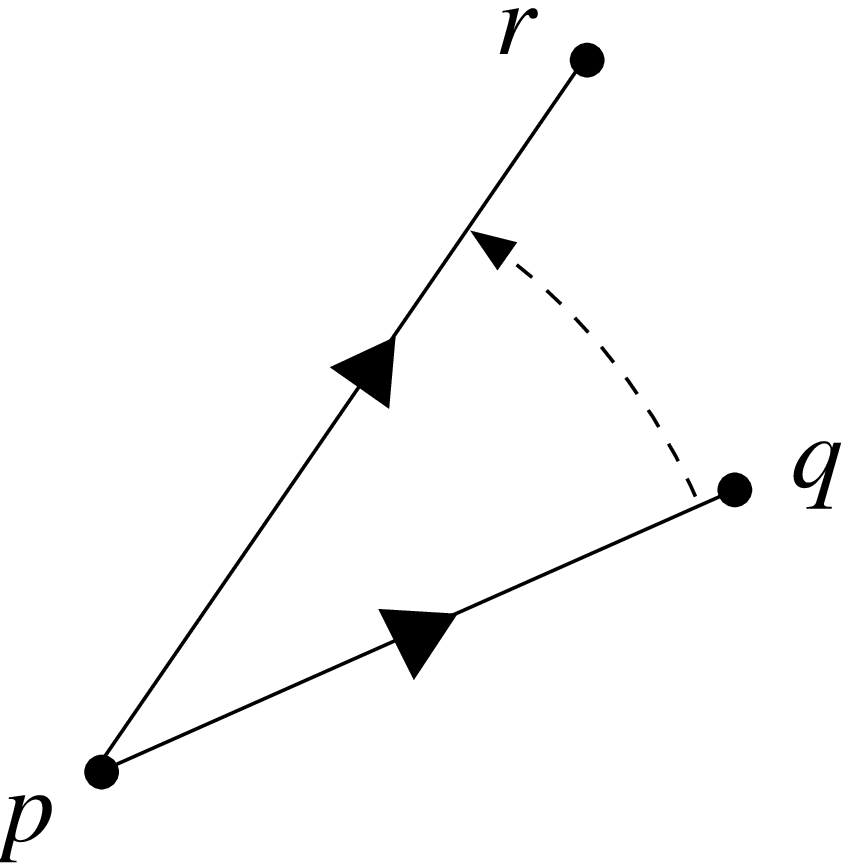}
\end{minipage}
\newpage
\nwenddocs{}\nwbegincode{20}\moddef{match and merge}\endmoddef\nwstartdeflinemarkup\nwenddeflinemarkup

        /***************************************
         * left_of ()
         ***************************************/

__device__ int left_of ( float2 r, float2 p, float2 q )
\{
  float value;

  value =
  (q.x - p.x) * (r.y - p.y) - (q.y - p.y) * ( r.x - p.x );

  return ( value > 0 );
\}

\nwendcode{}\nwbegindocs{21}\nwdocspar
Suppose $P$ and $Q$ are adjacent intervals of points
processed by a thread block in {\tt match\_and\_merge}.
Given two points $p$ and $q$
$q$ is either a corner of $H(Q)$ or is remote,
and $p$ is to the left of $Q$, there is a unique
tangent to $H(Q)$ from $P$: suppose $q'$ is the
corner of $H(Q)$ which supports the tangent.
Let $f(p,q)$ be {\tt LOW, EQUAL,} or {\tt HIGH}
according as $q$ is left of, at, or right of $q'$
(high if $q$ is remote).

Similarly if $p$ is remote or on $H(P)$ and $q$
is to the right of $P$, a function $f(p,q)$ indicates
whether $p$ is left of, at, or right of the
point supporting the tangent to $H(P)$ from $q$
(or remote).

These functions are implemented (on the device) by {\tt g} and {\tt f}
below, where $p$ = {\tt hood[i]} and $q$ = {\tt hood[j]}
and $P$ is defined by the range
{\tt start..start+d-1}, $Q$ by {\tt start+d..start+2*d-1}.

\bigskip

\centerline{\includegraphics[height=2in]{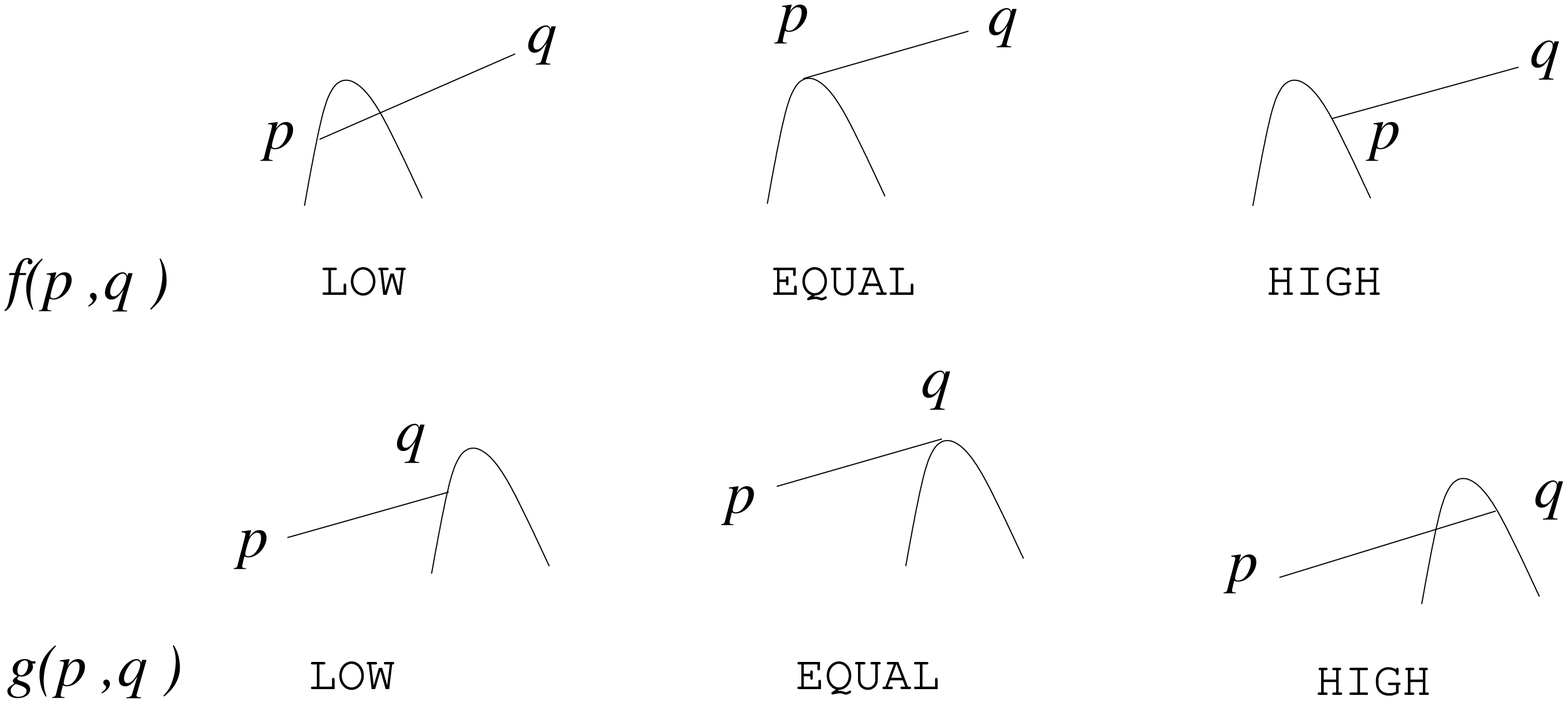}}

\nwenddocs{}\nwbegincode{22}\moddef{match and merge}\plusendmoddef\nwstartdeflinemarkup\nwenddeflinemarkup

#define LOW  -1
#define EQUAL 0
#define HIGH  1

__device__ short g( float2 * hood, short i, short j,
                      short start, short d )
\{
  float2 p, q, q_next, q_prev;
  int atstart, atend;
  int isleft;

  if ( hood [j] . x > 1 ) /* REMOTE */
    return HIGH;

  p = hood[i];
  q = hood[j];

  atend = ( j == start + 2*d - 1 || hood[j+1].x > 1.0 );

\nwendcode{}\nwbegindocs{23}\nwdocspar

{\tt Atend} signals the condition that $q$ is the rightmost
corner of $H(Q)$.  As written, it might cause thread divergence,
which could be remedied by adding an extra slot
in {\tt hood} and making it {\tt REMOTE.}
Using {\tt atend}, we can (without divergence)
make {\tt q\_next} default
to a point directly underneath the righmost corner
in $H(Q)$, in the case where $q$ is the last
corner in $H(Q)$.

If {\tt q\_next} is left of $p q$, then $q$
is {\tt LOW}.

\nwenddocs{}\nwbegincode{24}\moddef{match and merge}\plusendmoddef\nwstartdeflinemarkup\nwenddeflinemarkup
  
  q_next = hood [ j+1-atend ];
  q_next.y -= (float) atend;

  if ( left_of ( q_next, p, q ) )
        /*
         * avoidable divergence?
         */
    return LOW;

\nwendcode{}\nwbegindocs{25}\nwdocspar
Similarly {\tt atstart} indicates whether $q$ is
leftmost in $H(Q)$, in which case {\tt q\_prev} is
directly below it; otherwise it is the corner
of $H(Q)$ to its left;  $q$ is {\tt HIGH}
iff {\tt q\_prev} is left of the directed line-segment
$p q$.

\nwenddocs{}\nwbegincode{26}\moddef{match and merge}\plusendmoddef\nwstartdeflinemarkup\nwenddeflinemarkup

  atstart = ( j == start + d );
  q_prev = hood[ j + atstart - 1 ];
  q_prev.y -= (float) atstart;

  isleft = left_of ( q_prev, p, q );

  return HIGH * isleft + EQUAL * (1-isleft);
\}

        /*******************************
         *      f ( i, j, start, d )
         *******************************/

__device__ short f( float2 * hood, short i, short j,
                       short start, short d )
\{
  float2 p, q, p_next, p_prev;
  int atstart, atend;
  int isleft;

  if ( hood [i] . x > 1 ) /* REMOTE */
    return HIGH;

  p = hood[i];
  q = hood[j];

  atend = ( i == start + d - 1 || hood[i+1].x > 1 );
  
  p_next = hood [ i+1-atend ];
  p_next.y -= (float) atend;

  if ( left_of ( p_next, p, q ) )
    return LOW;

  atstart = ( i == start );
  p_prev = hood[ i + atstart - 1 ];
  p_prev.y -= (float) atstart;

  isleft = left_of ( p_prev, p, q );

  return HIGH * isleft + EQUAL * (1-isleft);
\}
\nwendcode{}\nwbegindocs{27}\nwdocspar
\begin{wrapfigure}{r}{2.7in}
\centerline{\includegraphics[width=2.5in]{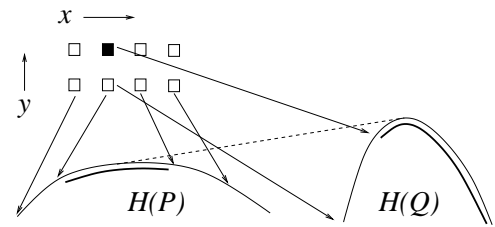}}
\caption{thread allocation.}
\label{fig: block}
\end{wrapfigure}

\noindent
{\sc The workhorse} of Wagener's algorithm is the
{\tt match\_and\_merge} procedure below.
Recall that $n/(2d)$ threads are launched
in blocks of dimension $d_1\times d_2$.  The
$\ell$-th block is to calculate $H(P\cup Q)$,
where $P$ and $Q$ are intervals of $d$ points
in {\tt hood} beginning at $2d\ell$ (this offset is
computed and stored in
{\tt start}). First {\tt start} and other
parameters are computed, and the scratch
array is set to a recognisably `uninitialised' value.
({\tt scratch[start..start+2*d-1]} is shared by the threads
in the same block).

The main effort is calculating the corners
of $H(P)$ and $H(Q)$ supporting the common
tangent.  Their indices will be placed
in {\tt pindex} and {\tt qindex}, initially
$-1$ to show uninitialised.

There are $d_1$ sample points along $H(P)$
and $d_2$ along $H(Q)$, but some of them
will be {\tt REMOTE}.

{\sc Match\_and\_merge} begins by setting the variables
$d_1, d_2, d,$ {\tt start}, $x,y,$, {\tt indx} to mirror
the construction of its thread blocks. Also,
{\tt pindex, qindex, scratch} are set to negative
values, meaning not initialised.
Also $i$ and $j$ are set to sample corners
(indices) in $H(P)$ and $H(Q)$.
There are $d_1$ sample indices $i$ and $d_2$ sample
indices $j$.  If $I$ is the set of sample indices $i$,
namely, $I = \{\text{\tt start} + d_2 x:~ 0 \leq x < d_1\}$,
and correspondingly $J = \{\text{\tt start} + d + d_1y:~ 0 \leq y < d_2\}$,
then the procedure is outlined as follows.

For $0 \leq x < d_1$, let $i_x = \text{\tt start}+d_2x$, so
$I = \{i_x:~ 0 \leq x < d_1\}$.  Also, for
$0 \leq y < d_2$, let $j_y = \text{\tt start} + d + d_1 y$,
so $J = \{ j_y: ~ 0 \leq y < d_2\}$.

\renewcommand{\L}{\textless}

\nwenddocs{}\nwbegincode{28}\moddef{match and merge}\plusendmoddef\nwstartdeflinemarkup\nwenddeflinemarkup
\LA{}mam 0: intialisations\RA{}
\LA{}mam 1: 0\L =x\L d1 scratch[start+x]=max jy g(ix,jy) \L = EQ\RA{}
\LA{}mam 2: 0\L =x\L d1 scratch[start+d+x]=j1(x)=unique j g(ix,j) =EQ\RA{}
\LA{}mam 3: scratch[start]=k0=max ix f(ix,j1(x)) \L = EQ\RA{}
\LA{}mam 4: 0\L =y\L d2 scratch[start+d+y]=ly=max jx, 0\L =x\L d2, g(k0+y,jx)\L =EQ\RA{}
\LA{}mam 5: scratch[start..]= unique p=k0+y,q=ly+x, g(p,q)=f(p,q)=EQ\RA{}
\LA{}mam 6: newhood[start..]= hood[start..p] catenated hood[q..start+d-1]\RA{}

\nwendcode{}\nwbegincode{29}\moddef{mam 0: intialisations}\endmoddef\nwstartdeflinemarkup\nwenddeflinemarkup
        /******************************
         * match_and_merge ()
         ******************************/

__global__ void match_and_merge ( float2 * hood, float2 * newhood,
        short * scratch )
\{
  int i, j, pindex, qindex, shift;
  int d1, d2, d, start, x, y, indx;

  d1 = blockDim.x;
  d2 = blockDim.y;
  d = d1 * d2;
  start = blockIdx.x * 2 * d;

  x = threadIdx.x;
  y = threadIdx.y;
  indx = x + d1 * y;

  pindex = qindex = -1;

  scratch[ start + indx     ] = -1;
  scratch[ start + indx + d ] = -1;
  
  __syncthreads();

  i = start + d2 * x;

\nwendcode{}\nwbegincode{30}\moddef{mam 1: 0\L =x\L d1 scratch[start+x]=max jy g(ix,jy) \L = EQ}\endmoddef\nwstartdeflinemarkup\nwenddeflinemarkup
  if ( hood[i].x <= 1.0 ) /* not REMOTE */
  \{
      j = start + d + d1 * y;

        /*
         * The condition below should identify the
         * unique interval of H(Q) touching the
         * tangent from hood[i].
         */

        if ( g(hood,i,j,start,d) <= EQUAL &&
           ( y == d2 - 1 ||
             hood[j+d1].x > 1.0 ||
             g(hood,i,j+d1,start,d) == HIGH )
         )
        scratch[ start+x ] = j;
  \}

  __syncthreads();

\nwendcode{}\nwbegincode{31}\moddef{mam 2: 0\L =x\L d1 scratch[start+d+x]=j1(x)=unique j g(ix,j) =EQ}\endmoddef\nwstartdeflinemarkup\nwenddeflinemarkup
  if ( hood[i].x <= 1.0 )
  \{
    j = scratch[start + x] + y;

    if ( g(hood,i,j,start,d) == EQUAL )
      scratch[start + d  + x] = j;
    else if ( d2 < d1 && g(hood,i,j+d2,start,d) == EQUAL )
      scratch[start + d + x] = j+d2;
  \}

  __syncthreads();
\nwendcode{}\nwbegindocs{32}\nwdocspar
{\sc Suppose that} $p$ and $q$ are the actual
corners to be calculated, supporting the common tangent to
$H(P)$ and $H(Q)$.
For each sample point $p_i$ a corresponding
tangent corner $q_i'$ on $H(Q)$ has been calculated.

\begin{theorem}
The tangent corners $q_i'$ occur in nondecreasing
left-to-right order, and $p_i$ is left of, equal to,
or right of $p$ according as $f(p_i,q_i')$ is
{\tt LOW,} {\tt EQUAL,} or {\tt HIGH}.
\end{theorem}

{\bf Sketch proof.}  Parametrise the tangents
to $H(Q)$ by the angle $\theta$ they make with the
$x$-axis:
$\theta$ varies over the clockwise interval from $90^\circ$
(yielding the left vertical tangent) to $-90^\circ$.

\begin{wrapfigure}{r}{2.7in}
\centerline{\includegraphics[width=2.5in]{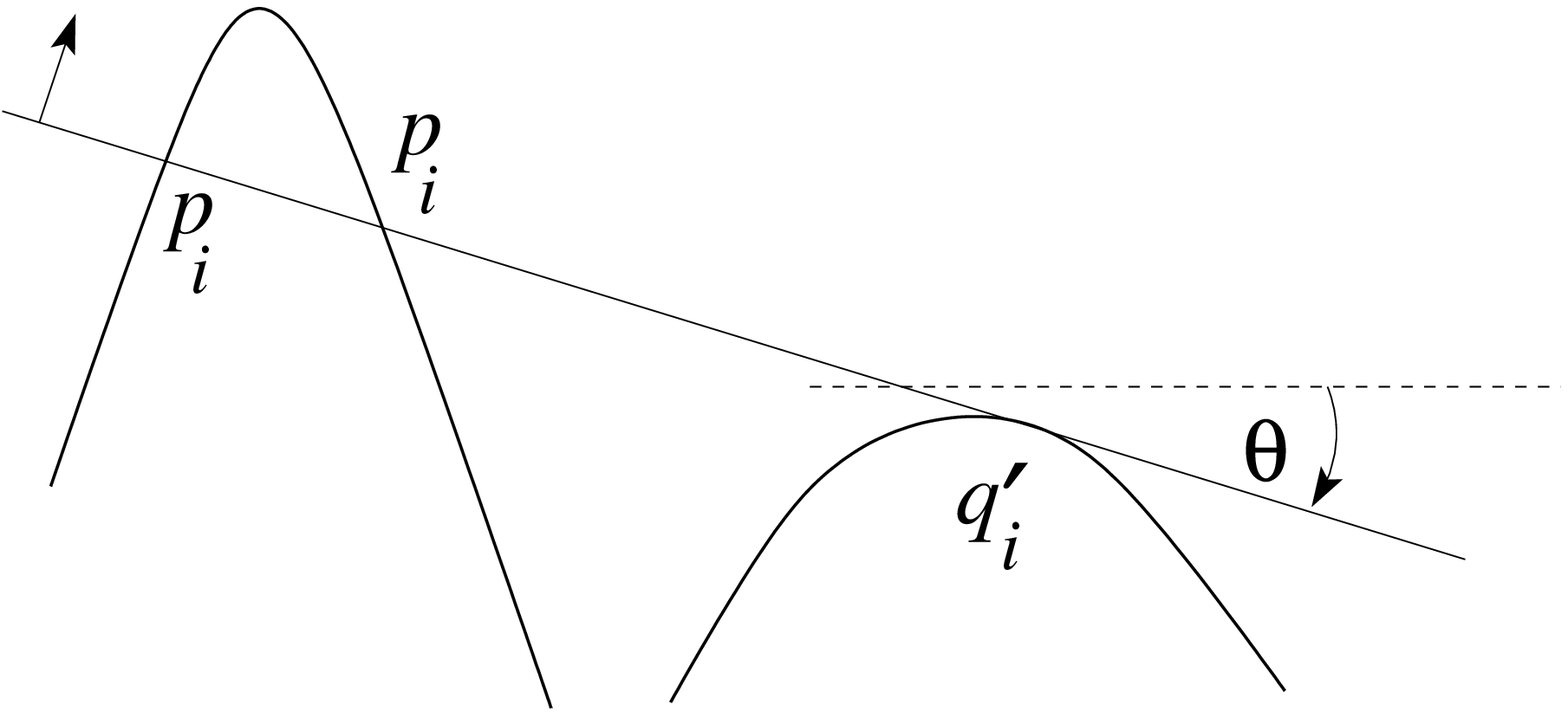}}
\caption{$L_\theta$.}
\label{fig: monot}
\end{wrapfigure}

For each $\theta$, let $L_\theta$ be the half-plane
left of the tangent line at angle $\theta$
(except at $\pm 90^\circ$, this means above the tangent line).
The map $\theta\to L_\theta$ is, loosely speaking,
continuous, and $H(P)\cap L_\theta$ contracts
with $\theta$.  The point of contact between
$L_\theta$ and $H(Q)$ shifts discontinuously from
corner to corner, but always rightward.
At a unique angle, $\theta=\alpha$, say, the
intersection contains a single point, and that
point is $p$.  The points $p_i$ under consideration
are left and right endpoints of various sets
$H(P)\cap L_\theta$, the points $q'_i$
are points of contact between various $L_\theta$
and $H(Q)$, and the points $p_i$ are left of,
at, or right of $p$ according to
the values of $f(p_i,q'_i)$.\qed

\nwenddocs{}\nwbegincode{33}\moddef{mam 3: scratch[start]=k0=max ix f(ix,j1(x)) \L = EQ}\endmoddef\nwstartdeflinemarkup\nwenddeflinemarkup
  j = scratch[start+d+x];
  if ( hood[i].x <= 1.0 &&
         f(hood,i,j,start,d) <= EQUAL &&
         ( x == d1-1 ||
           hood[i+d2].x > 1.0 ||
           f(hood,i+d2,scratch[start + d + x + 1],start,d) == HIGH
         )
     )
  scratch[start] = i;

  __syncthreads();

\nwendcode{}\nwbegincode{34}\moddef{mam 4: 0\L =y\L d2 scratch[start+d+y]=ly=max jx, 0\L =x\L d2, g(k0+y,jx)\L =EQ}\endmoddef\nwstartdeflinemarkup\nwenddeflinemarkup
  i = scratch[start] + y;

  if ( hood[i].x <= 1.0 ) /* not REMOTE */
  \{
    j = start + d + x * d2;
    if ( g(hood,i,j,start,d) <= EQUAL &&
      ( x == d1 - 1 ||
        hood[j+d2].x > 1.0 ||
        g(hood,i,j+d2,start,d) == HIGH )
      )
      scratch[start + d + y] = j;
  \}

  __syncthreads();

\nwendcode{}\nwbegincode{35}\moddef{mam 5: scratch[start..]= unique p=k0+y,q=ly+x, g(p,q)=f(p,q)=EQ}\endmoddef\nwstartdeflinemarkup\nwenddeflinemarkup
  j = scratch[ start + d + y ] + x;

    if ( x < d2 &&
        g(hood,i,j,start,d) == EQUAL
        &&
        f(hood,i,j,start,d) == EQUAL )
    \{
      scratch [start    ] = i;
      scratch [start + 1] = j;
    \}

  __syncthreads();

\nwendcode{}\nwbegincode{36}\moddef{mam 6: newhood[start..]= hood[start..p] catenated hood[q..start+d-1]}\endmoddef\nwstartdeflinemarkup\nwenddeflinemarkup
  pindex = scratch [ start     ];
  qindex = scratch [ start + 1 ];

  newhood [ start + indx ] = hood[ start + indx ];
  make_remote ( & ( newhood [ start + d + indx ] ) );
  __syncthreads();

\nwendcode{}\nwbegindocs{37}\nwdocspar
Let $s$ be the 'shift', {\tt qindex-pindex-1.}\hfil\break
Then {\tt hood[qindex\ldots start+2*d-1]} is copied, shifted
left by $s$, to\hfil\break
{\tt newhood[pindex+1\ldots]}.

\nwenddocs{}\nwbegincode{38}\moddef{mam 6: newhood[start..]= hood[start..p] catenated hood[q..start+d-1]}\plusendmoddef\nwstartdeflinemarkup\nwenddeflinemarkup
  shift = qindex - pindex - 1;

  if ( start + d + indx >= qindex )
    newhood [ start + d + indx - shift ] = hood [ start + d + indx ];

  __syncthreads();

\nwendcode{}\nwbegindocs{39}\nwdocspar
Final closing brace in match and merge.
\nwenddocs{}\nwbegincode{40}\moddef{match and merge}\plusendmoddef\nwstartdeflinemarkup\nwenddeflinemarkup
\}

\nwendcode{}\nwbegincode{41}\moddef{wagener.Makefile}\endmoddef\nwstartdeflinemarkup\nwenddeflinemarkup
.SUFFIXES: .nw .tex .c

wagener: wagener.nw
        /usr/bin/notangle -Rwagener -L wagener.nw > wagener.cu

dvi: wagener.nw
        /usr/bin/noweave -delay wagener.nw > wagener.tex
        latex wagener
        latex wagener
        rm wagener.out wagener.aux

clean:
        rm *.c *.dvi *.log
\nwendcode{}\nwbegindocs{42}\nwdocspar
\begin{figure}
\centerline{\includegraphics[width=3in]{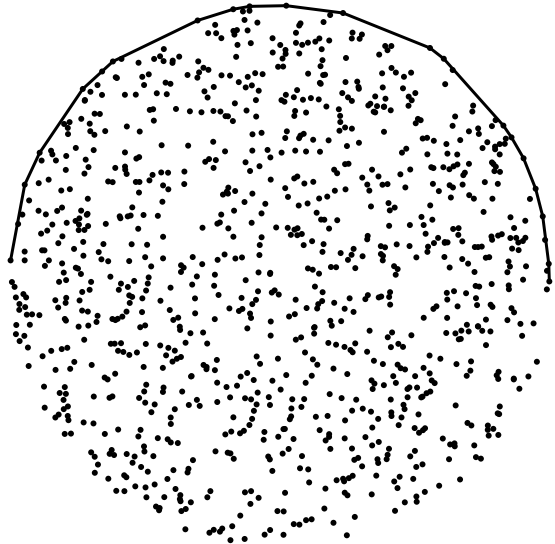}}
\caption{Sample cuda output, 1024 points}
\label{fig: cuda1024}
\end{figure}

\section{Conclusions}

Wagener's PRAM algorithm, published only as a manuscript,
is very clean and simple in comparison, for example, with
another $O(\log n)$ algorithm in [\ref{acgoy}].

Our program illustrates how Wagener's PRAM algorithm might be realised
on a CUDA chip: the organisation, at any rate, is faithful to
the model.  However, it is insensitive to the memory
bank conflicts which make the chip, although robust enough
to tolerate these conflicts, so slow that the parallel program
is slow by comparison with another serial program (not described here).

On the other hand, we tried to avoid branching, another reason
for serialisation, and the writing of branch-free code is an
interesting challenge.

Another possible innovation was our usage of padding, rather
than compression, which we felt too cumbersome.  That is,
data would be in blocks, with `live' data to the left of the block
padded with `remote' values  on the right.  This left some threads
with nothing to do, but it avoided allocation tasks.

A few last words about optimal speedup.  Our algorithm gets
the data points in sorted order, and in principle should use
$O(n)$ work (runtime $\times$ processor count): but it uses
$O(n\log n)$.  We indicate how Wagener's algorithm can achieve optimal
speedup:
$O(\log n)$ time and $O(n)$ work.  So we suppose we have $n$
data points and $n/(\log_2 n)$ processors.

\begin{itemize}
\item
Separate the data into $n/\log n$ strips, 1 per processor,
and compute the convex hood in each strip, $O(\log n)$ time
serially.
\item
Store the hood corners in each strip (in left-to-right order)
in balanced trees of size $\leq \log n$.
\item
Overmars and Van Leeuwen devised a logarithmic time
procedure, a balanced search, for locating common tangents:
see [\ref{acgoy},\ref{od}].
Applying their procedure to convex
hoods stored in balanced trees, convex hoods can be merged in
logarithmic time.
\item
This means that with $\log \log n$ passes using $\leq n/\log n$
processors per pass, convex hoods can be calculated for
$n/\log^2 n$ strips each containing $\log^2 n$ points,
each in time $O(\log\log n)$, hence $O((\log \log n)^2)$ overall,
which is of course $O(\log n)$.
\item
Under the PRAM model, these trees can be flattened
into arrays using $\log n$ processors per tree.
Now we have the same organisation as in our Cuda
algorithm, with strips of $\log^2 n$ points each
stored in an array.
\item
Our implementation involved
finding the common tangent between adjacent hoods
using $k$ processors for hoods of size (at most) $k$,
in $O(1)$ time.

Given $k \geq \log^2 n$, this
can
be done with $k/\log n$ processors.  In this
case there are at least $\sqrt{k}$ processors available.
Let $h = \sqrt[4]{k}$, and let $P$ be the points in
the left-hand strip and $Q$ the points on the right.
Subdivide $H(P)$ into $k/h$ intervals of length $h$.
For each interval endpoint $p$, allocate $h$ processors
which first inspect intervals in $H(Q)$ of length
$k/h$, bracketing the tangent from $p$ to one of
these intervals; next they bracket the tangent to
an interval of length $k/h^2$, then $k/h^3$, and
finally return the tangent from $p$ to $H(Q)$.
This brackets the common tangent endpoint in $H(P)$
to an interval of length $k/h$; repeat the process
to bracket to intervals of length $k/h^2$ and $k/h^3$, and
finally compute the common tangent.
\end{itemize}

When run on the dataset illustrated, our CUDA algorithm
is perceptibly slower by comparison with a serial algorithm
(which is not described here).  This is not surprising
considering the serialisation of conflicting memory accesses.
To attempt optimal speedup as described here would
demand a great deal of effort.  Our CUDA program is a specimen
implementation of a PRAM algorithm which cannot claim much
speed advantage.

\section{References}

\begin{enumerate}
\item
\label{acgoy}
`Parallel Computational Geometry,' with Alok Aggarwal, Bernard Chazelle,
Leo Guibas, and Chee-Keng Yap (1988).
{\em Algorithmica \bf 3}, 293-327  (special issue on Parallel Processing).

\item          
\label{od}
\lq Some parallel geometric algorithms' (1993), in
{\em Lectures on Parallel Computation,} ed.\ Alan Gibbons
and Paul Spirakis, Cambridge University Press,
77--108.

\item
\label{wagener}
H. Wagener (1985). Optimally parallel algorithms for
convex hull determination.  Manuscript, Technical University
of Berlin.
\end{enumerate}

\end{document}